\pdfoutput=1
\documentclass{article}
\usepackage[utf8]{inputenc}
\usepackage{authblk}
\usepackage{url}
\usepackage{biblatex}
\usepackage{amsmath}
\usepackage{subfig}
\usepackage{appendix}
\usepackage{enumitem}
\usepackage{arxiv}
\usepackage{booktabs}
\title{\textbf{Designing Refund Bonus Schemes for Provision Point Mechanism in Civic Crowdfunding}}
\author[1]{Sankarshan Damle\thanks{sankarshan.damle@research.iiit.ac.in}}
\author[2]{Moin Hussain Moti\thanks{mhmoti@cse.ust.hk}}
\author[1]{Sujit Gujar\thanks{sujit.gujar@iiit.ac.in}}
\author[3]{Praphul Chandra\thanks{praphulcs@koinearth.com}}
\affil[1]{Machine Learning Lab, International Institute of Information Technology (IIIT), Hyderabad}
\affil[2]{The Hong Kong University of Science and Technology (HKUST), Hong Kong}
\affil[3]{KoineArth Inc., Bangalore}
\date{}

\usepackage[export]{adjustbox}
\usepackage{comment}
\usepackage{float}
\usepackage[linesnumbered,ruled,vlined]{algorithm2e}
\usepackage{dirtytalk}
\usepackage{verbatim}
\usepackage{amssymb}
\usepackage{soul}
\usepackage{amsmath}
\usepackage{dirtytalk}
\usepackage{multirow}
\usepackage{amsthm}

\usepackage{blindtext,graphicx}
\usepackage[absolute]{textpos}
\newtheorem{condition}{Condition}
\newtheorem*{claim}{Claim}
\newtheorem{Claim}{Claim}
\newtheorem{Property}{Property}
\newtheorem*{Property*}{Property}
\newtheorem{corollary}{Corollary}
\newtheorem{proposition}{Proposition}
\newtheorem{definition}{Definition}

\newtheorem*{theorem}{Theorem}

\newtheorem{innercustomthm}{Theorem}
\newenvironment{customthm}[1]
{\renewcommand\theinnercustomthm{#1}\innercustomthm}
{\endinnercustomthm}

\newcommand{\squishlist}{
 \begin{list}{$\bullet$}
  { \setlength{\itemsep}{8pt}
     \setlength{\parsep}{3pt}
     \setlength{\topsep}{3pt}
     \setlength{\partopsep}{0pt}
     \setlength{\leftmargin}{0.025cm}
     \setlength{\labelwidth}{1em}
     \setlength{\labelsep}{0.5em} } }
\newcommand{\squishend}{
  \end{list}  }

\begin{document}

\maketitle

\begin{textblock}{15}(3.15,1)
\centering
\noindent\scriptsize A version of this paper appears in the Proceedings of the 18th Pacific Rim International Conference on Artificial Intelligence \textit{(PRICAI-2021)} and the 2nd Games, Agents, and Incentives Workshop (GAIW) held at \textit{AAMAS 2020}. This is a full version.
\end{textblock}

\begin{abstract}
Civic crowdfunding (CC) is a popular medium for raising funds for civic projects from interested agents. With Blockchains gaining traction, we can implement CC in a reliable, transparent, and secure manner with smart contracts (SCs). The fundamental challenge in CC is free-riding. PPR, the proposal by Zubrickas \cite{zubrickas2014provision} of giving refund bonus to the contributors, in the case of the project not getting provisioned, has attractive properties. However, as observed by Chandra et al. \cite{chandra2016crowdfunding}, PPR faces a challenge wherein the agents defer their contribution until the deadline. We define this delaying of contributions as a race condition. To address this, their proposal, PPS, considers the temporal aspects of a contribution. However, PPS is computationally complex,  expensive to implement as an SC, and it being sophisticated, it is difficult to explain to a layperson. In this work, our goal is to identify all essential properties a refund bonus scheme must satisfy in order to curb free-riding while avoiding the race condition. We prove Contribution Monotonicity and Time Monotonicity are sufficient conditions for this. We propose three elegant refund bonus schemes satisfying these two conditions leading to three novel mechanisms for CC - PPRG, PPRE, and PPRP. We show that PPRG is the most cost-effective mechanism when deployed as an SC. We show that under certain modest assumptions on valuations of the agents, in PPRG, the project is funded at equilibrium.
\end{abstract}

\section{Introduction}

\emph{Crowdfunding} is the practice of raising funds for a project through voluntary contributions from a large pool of interested participants and is an active research area \cite{Alaei:2016:DMC:2940716.2940777,DBLP:conf/wine/2017,chandraAamas17,shen2018information,strausz2017theory}. For \textit{private} projects, specific reward schemes incentivize the participants to contribute towards crowdfunding. Using crowdfunding to raise funds for \textit{public} (non-excludable) projects\footnote{For example, the crowdfunding of the Wooden Pedestrian Bridge in Rotterdam: \url{https://www.archdaily.com/770488/the-luchtsingel-zus}}, however, introduces the free-riding problem. Observe that we cannot exclude non-contributing participants from enjoying the benefits of the public project. Thus, strategic participants, henceforth \emph{agents}, may not contribute. If we can address this challenge, ``civic" crowdfunding (CC), i.e., crowdfunding of public projects, can lead to greater democratic participation. It also contributes to citizens' empowerment by increasing their well-being by solving societal issues collectively. Thus, this paper focuses on solving the challenge of free-riding in CC.

With the advancement of the \emph{blockchain} technology,   \emph{smart contracts} (SC) now allow for the deployment of such CC projects. A smart contract is a computer protocol intended to digitally facilitate, verify, or enforce the negotiation or performance of a contract \cite{wiki:SmartContract}. Since a crowdfunding project as an SC is on a trusted, publicly distributed ledger, it is open and auditable. This property makes the agents' contributions and the execution of the payments transparent and anonymous. Besides, there is no need for any centralized, trusted third party, which reduces the cost of setting up the project. WeiFund \cite{weifund} and Starbase \cite{starbase} are examples of decentralized crowdfunding platforms on public blockchains like \emph{Ethereum}. 

Traditionally, a social planner uses the voluntary contribution mechanism with a provision point, \emph{provision point mechanism} (PPM) \cite{bagnoli1989provision}. The social planner sets up a target amount, namely the provision point. If the net contribution by the agents crosses this point, the social planner executes the project. We call this as \emph{provisioning} of the project. Likewise, the project is said to be \emph{under-provisioned}, if the net contribution does not exceed the provision point.  In the case of under-provisioning, the planner returns the contributions. PPM has a long history of applications, but consists of several inefficient equilibria \cite{bagnoli1989provision,schmidtz1991limits}.

Zubrickas proposes \textit{Provision Point mechanism with Refund} (PPR), which introduces an additional \emph{refund bonus} to be paid to the contributing agents. This refund is paid along with each agent's contribution, in the case of under-provisioning of the project \cite{zubrickas2014provision}. This incentive induces a simultaneous move game in PPR, in which the project is provisioned at equilibrium. Chandra et al.~\cite{chandra2016crowdfunding} observe that PPR may fail in online settings (e.g., Internet-based platforms~\cite{wiki:gofundme,wiki:kickstarter}) since, in such a setting, an agent can observe the current amount of funds raised. Hence, in online settings, strategic agents in PPR would choose to defer their contributions until the end to check the possibility of free-riding and contribute only in anticipation of a refund bonus. Such deference leads to a scenario where every strategic agent competes for a refund bonus at the deadline. We refer to this scenario as a \emph{race condition}. As the agents can observe the contributions' history in online settings, it induces a sequential game. Thus, we refer to such settings as \emph{sequential settings}.

\textit{Provision Point mechanism with Securities} (PPS)~\cite{chandra2016crowdfunding} introduces a class of mechanisms using complex prediction markets \cite{abernethy2013efficient}. These markets incentivize an agent to contribute \textit{as soon as it arrives}, thus avoiding the race condition. The challenge with the practical implementation of sophisticated mechanisms such as PPS is that as it uses complex prediction markets, it is difficult to explain to a layperson and computationally expensive to implement, primarily as an SC. 

The introduction of the refund bonus is \textit{vital} in these mechanisms as it incentivizes agents to contribute, thus avoiding free-riding. Consequently, we focus on provision point mechanisms with a refund bonus. Our primary goal is to abstract out conditions that \emph{refund bonus schemes} should satisfy to avoid free-riding and the race condition. 
We believe that such a characterization would further make it easier to explore simpler and computationally efficient CC mechanisms.

Towards this, we introduce, \emph{Contribution Monotonicity} (CM) and \emph{Time Monotonicity} (TM). Contribution monotonicity states that an agent's refund should increase with an increase in its contribution. Further, time monotonicity states that an agent's refund should decrease if it delays its contribution. We prove these two conditions are \emph{sufficient} to provision a public project via crowdfunding in a sequential setting at equilibrium and avoid the race condition (Theorem \ref{Theorem:General}). We also prove that TM and \textit{weak} CM are also \textit{necessary}, under certain assumptions on equilibrium behavior (Theorem \ref{thm::neccGeneral}).

With these theoretical results on CM and TM, we propose three elegant refund bonus schemes that satisfy CM and TM. These schemes are straightforward to explain to a layperson and are computationally efficient to implement as an SC. With these three schemes, we design novel mechanisms for CC, namely \emph{Provision Point mechanism with Refund through Geometric Progression} (PPRG); \emph{Provision Point mechanism with Refund based on Exponential function} (PPRE), and \emph{Provision Point mechanism with Refund based on Polynomial function} (PPRP). We analyze the cost-effectiveness of these mechanisms and PPS when deployed as SCs and show that PPRG is significantly more cost-effective, i.e., PPRG requires the least amount of capital to set up.

\section{Preliminaries}
\label{sec:prelim}
We focus on Civic Crowdfunding (CC) which involves provisioning of projects without coercion where agents arrive over time and \emph{not} simultaneously, i.e., CC in a sequential setting. We assume that agents are aware of the history of contributions, i.e., the provision point and the total amount remaining towards the project's provision at any time. However, the agents have no information regarding the number of agents yet to arrive or the agents' sequence. Ours is the first attempt at providing a general theory for refund bonuses in CC to the best of our knowledge. Thus, we also assume that agents do not have any other information regarding the project. This information can be arbitrarily anything. E.g., an agent may deviate from its strategy if it knows about spiteful contributions and related corruption. Thus, unlike \cite{damle2019aggregating,damleijcai19}, every agent's belief is symmetric towards the project's provision~\cite{bagnoli1989provision,chandra2016crowdfunding,zubrickas2014provision}.

\subsection{Model}
     
A social planner (SP) proposes crowdfunding a public project $P$ on a web-based crowdfunding platform; we are dealing with sequential settings.  SP seeks voluntary contributions towards it. The proposal specifies a target amount $H$ necessary for the project to be provisioned, referred to as the \emph{provision point}. It also specifies deadline $(T)$ by which the funds need to be raised. If the target amount is not achieved by the deadline, the project is not provisioned, i.e., the project is under-provisioned. In the case of under-provisioning, the SP returns the contributions.

A set of agents $N =\{1,2,\dots,n\}$ are interested in the crowdfunding of $P$. An Agent $i \in N$ has value $\theta_i \geq 0$ if the project is provisioned. It arrives at time $y_i$ to the project, observes its valuation $(\theta_i)$ for it \emph{as well as} the net contribution till $y_i$. However, no agent has knowledge about any other agent's arrival or their contributions towards the project. 

Agent $i$ may decide to contribute $x_i \geq 0$ at time $t_i$, such that $y_i\leq t_i\leq T$, towards its provision.  Let $\vartheta = \sum_{i=1}^{i=n} \theta_i$ be the total valuation, and $C=\sum_{i=1}^{i=n} x_i$ be the sum of the contributions for the project. We denote $h^t$ as the amount that remains to be funded at time $t$.

 A project is provisioned if $C\geq H$ and under-provisioned if $C < H$, at the end of deadline $T$. SP keeps a budget $B$ aside to be distributed as a refund bonus among the contributors if the project is under-provisioned. This setup induces a game among the agents as the agents may now contribute to getting a fraction of the budget $B$ in anticipation that the project may be under-provisioned.

 Towards this, let $\sigma = (\sigma_1,\dots,\sigma_n)$ be the vector of strategy profile of every agent where Agent $i$'s strategy consists of the tuple $\sigma_i = (x_i, t_i)$, such that $x_i \in [0,\theta_i] $ is its voluntary contribution to the project at time $t_i \in [y_i,T]$. We use the subscript $-i$ to represent vectors without Agent $i$. The payoff for an Agent $i$ with valuation $\theta_i$ for the project, when all the agents play the strategy profile $\sigma$ is $\pi_i(\sigma;\theta_i)$. Note that, in this work, we assume that every agent only contributes once to the project. We justify this assumption while providing the strategies for the agents (Section \ref{sec::PPRG}). We leave it for future study to explore the effect of splitting of an agent's contribution to the project's provision and its payoff.
    
 Let $\mathcal{I}_X$ be an indicator random variable that takes the value 1 if $X$ is true and 0 otherwise. Further, let $R:\sigma\rightarrow\mathbf{R}^n$ denote the refund bonus scheme. Then the payoff structure for a provision point mechanism with a refund bonus scheme $R(\cdot)$ and budget $B$, for every Agent $i$ contributing $x_i$ and at time $t_i$, will be
    \begin{equation}
    	\label{GeneralPayoff}
		\pi_i(\sigma;\theta_i)=\mathcal{I}_{C\geq H}(\theta_i-x_i)+\mathcal{I}_{C<H}\left(R_i(\sigma)\right),
    \end{equation}
    where $R_i(\sigma)$ is the share of refund bonus for Agent $i$ as per $R(\sigma)$ such that $R(\sigma)=\left(R_1(\sigma),\dots,R_n(\sigma)\right)$. We use $R(\cdot)$ to denote a refund bonus scheme and $R_i(\cdot)$ to denote Agent $i$'s share of the refund bonus as per $R(\cdot)$ whenever the inputs are obvious.
    
    \smallskip
    \noindent\textit{Important Game-Theoretic Definitions.}
We require the following definitions for the understanding of the results presented in this paper.

        	\begin{definition}[Pure Strategy Nash Equilibrium (PSNE)]
        	A strategy profile $\sigma^* = (\sigma_1^*,\dots,\sigma_n^*)$ is said to be a Pure Strategy Nash equilibrium (PSNE) if for every Agent $i$, it maximizes the payoff $\pi_i(\sigma^*;\theta_i)$ i.e., $\forall i \in N$, $$\pi_i(\sigma_i^*,\sigma^*_{-i};\theta_i) \geq \pi_i(\sigma_i,\sigma^*_{-i};\theta_i) \ \forall \sigma_i, \forall \theta_i.$$
        	\end{definition}
The strategy profile for the Nash Equilibrium is helpful in a simultaneous move game. However, for sequential settings, where the agents can see the actions of the other agents, they may not find it best to follow the PSNE strategy. For this, we require a strategy profile that is the best response of every agent during the project, i.e., the best response for every sub-game induced during it. Such a strategy profile is said to be a \emph{Sub-game Perfect Equilibrium}.
            \begin{definition}[Sub-game Perfect Equilibrium (SPE)] A strategy profile $\sigma^* = (\sigma_1^*,\dots,\sigma_n^*)$, with $\sigma_i^*=(x_i^*,t_i^*)$, is said to be a sub-game perfect equilibrium if for every Agent $i$, it maximizes the payoff $\pi_i(\sigma^*_i,\sigma^*_{-i|H^{t_i^*}};\theta_i)$ i.e. $\forall i \in N$,
            $$\pi_i(\sigma^*_i,\sigma^*_{-i|H^{t_i^*}};\theta_i) \geq \pi_i(\sigma_i,\sigma^*_{-i|H^{t_i^*}};\theta_i) \ \forall\sigma_i, \forall H^{t}, \forall \theta_i.$$ 
            \end{definition}
   
       Here, $H^t$ is the history of the game till time $t$, constituting the agents' arrivals and their contributions and $\sigma^*_{-i|H^{t_i^*}}$ indicates that the agents who arrive after $t_i^*$ follow the strategy specified by $\sigma^*_{-i}$. Informally, at every stage of the game, it is Nash Equilibrium for each agent to follow the SPE strategy irrespective of what has happened.

In this work, we aim to derive deterministic strategies for the induced CC game. Non-deterministic strategies in our context will refer to equilibrium concepts like Bayesian Nash equilibrium (BNE). A layperson will be required to perform complex randomization to play such a strategy in practice. Besides, it will also need assurance over the correctness of its calculation. As a result, we focus on PSNE, a more robust and straightforward notion a layperson to play in practice. The choice of PSNE is also consistent with the CC literature.

\section{Related Work}
This paper focuses on the class of mechanisms that require the project to aggregate a minimum level, provision point, of funding before the SP can claim it. There is extensive literature on mechanism design for CC with provision point (see \cite{chandra2016crowdfunding} and the references therein). Our work is most closely related to PPM, PPR, and PPS.

\smallskip
\noindent\textit{Provision Point Mechanism (PPM).}
PPM \cite{bagnoli1989provision} is the simplest mechanism in this class where agents contribute voluntarily. Agents gain a positive payoff only when the project gets provisioned and a payoff of zero otherwise i.e., $R^{PPM}(\sigma)=\left((0)\:|\: \forall i \in N\right)$. Then the payoff structure of PPM, for every Agent $i$, is,
\begin{equation*}
	\pi_i(\cdot) = \mathcal{I}_{C \geq H} \times (\theta_i - x_i) 
\end{equation*}
where, $\pi_i(\cdot) ~ \mbox{and} ~ x_i$ are Agent $i$'s payoff and contribution respectively. PPM has been shown to have multiple equilibria and also does not guarantee strictly positive payoff to the agents. It has led the mechanism to report under-provisioning of the project, i.e., the provision point not being reached.

\smallskip
\noindent\textit{Provision Point Mechanism With Refund (PPR).}
 PPR \cite{zubrickas2014provision} improves upon the limitations of PPM by offering refund bonuses to the agents in case the project does not get provisioned. This refund bonus scheme is directly proportional to agent's contribution and is given as $R^{PPR}_i(\sigma)=\left(\frac{x_i}{C}\right)B \ \forall i \in N$, where $B>0$ is the total budget. Then the payoff structure of PPR, for every Agent $i$ is,
\begin{equation*}
	\pi_i(\cdot) = \mathcal{I}_{C \geq H} \times (\theta_i - x_i) + \mathcal{I}_{C < H} \times R^{PPR}_i(\sigma).
\end{equation*}

In PPR, an agent does not know other agents' contributions. 
Thus, as shown in \cite{chandra2016crowdfunding}, PPR collapses to a one-shot simultaneous game where every agent delays its contribution till the deadline. This delay results in each agent attempting to contribute at the deadline, leading to a \emph{race condition}, defined as follows.

	\begin{definition}[Race Condition]\label{def:race_cond} A strategy profile $\sigma^* = (\sigma_1^*,\dots,\sigma_n^*)$ is said to have a race condition if $\exists S\subseteq N\mbox{~with~}|S|>1$, for which $\forall i\in S$ the strategy 
	$\sigma_i^*=(x_i^*,t)$, with $x_i^*$ as the equilibrium contribution, is the PSNE of the induced game i.e., $\forall 
	\sigma_i,\forall i \in S$, 
            $$
            \pi_i(\sigma_i^*,\sigma^*_{-i};\theta_i) \geq \pi_i(\sigma_i,\sigma^*_{-i};\theta_i) \mbox{~where~} t\in[\bar{y},T]\mbox{~s.t.~} \bar{y}= \max_{j \in S} y_j.
            $$ 
            \end{definition}
            Here, $\sigma_i=(x_i^*,t_i)$ $\forall t_i \in [y_i,T]$.
            
            For PPR, $S=N$ and $t=T$, i.e., the strategy $\sigma_i^*=(x_i^*,T) \ \forall i\in N$ constitutes a set of PSNE of PPR in a sequential setting as the refund bonuses here are independent of time of contribution. Thus, agents have no incentive to contribute early. Such strategies lead to the project not getting provisioned in practice and are undesirable. 

\smallskip
\noindent\textit{Provision Point Mechanism With Securities (PPS).}
PPS \cite{chandra2016crowdfunding} addresses the shortcomings of PPR by offering early contributors higher refund than a late contributor for the same amount. The refund bonus of a contributor is determined using securities from a cost based complex prediction market \cite{abernethy2013efficient} and is given as $R^{PPS}_i(\sigma)= (r^{t_i}_i - x_i) \ \forall i \in N$ where, $t_i ~ \mbox{and} ~ r_i^{t_i}$ are Agent $i$'s time of contribution and the number of securities allocated to it, respectively. $r_i^{t_i}$ depends on the contribution $x_i$ and the total number of securities issued in the market at the time contribution $t_i$ denoted by $q^{t_i}$. Then the payoff structure of PPS, for every Agent $i$, can be expressed as,
\begin{equation*}
	\pi_i(\cdot) = \mathcal{I}_{C \geq H} \times (\theta_i - x_i) + \mathcal{I}_{C < H} \times R^{PPS}_i(\sigma)
\end{equation*}

To set up a complex prediction market in the context of CC, PPS requires a cost function ($C_0$) satisfying \cite[CONDITIONS 1-4,6-7]{chandra2016crowdfunding}.
$C_0$ can either be based on the \emph{logarithmic}~\cite[Eq. 3]{chandra2016crowdfunding} or the \emph{quadratic} scoring rule~\cite[Eq. 4]{chandra2016crowdfunding}.

PPS awards every contributing agent securities for the project not getting provisioned. These securities are dependent on the agent contribution, i.e., the greater the contribution, the higher the number of securities are allocated to the agent. Each of these securities pays out a unit amount if the project is not provisioned. However, setting up such a market and computing securities to be allotted is computationally expensive to implement as a smart contract. Hence, we want to look for more desirable refund bonus schemes.

\section{Desirable Properties of Refund Bonus Schemes}

Motivated by the theoretical guarantees of PPR and PPS, we look for CC mechanisms with refund bonus schemes in this paper. In this context, a \emph{desirable} refund bonus scheme should not just restrict the set of strategies so that the project is provisioned at equilibrium, but should also incentivize \emph{greater} and \emph{early} contributions, to avoid the race condition, from all interested agents. A refund bonus scheme without these would fail in a sequential (web-based) setting, similar to PPR, and hence these are essential for a provision point mechanism's implementation online. 
We formalize these desirable properties as the following two \emph{conditions} for a refund bonus scheme $R(\sigma)$ where $\sigma=\left((x_i,t_i) \: | \: \forall i \in N \right)$ such that $x_i \in (0,H]$, $t_i \in [y_i,T]$ $\forall i \in N$ and with budget $B$. 

	\begin{condition}[Contribution Monotonicity]
		\label{Condition:1}
    The refund must always increase with the increase in contribution so as to incentivize greater contribution i.e., $\forall i\in N,$ $R_i(\sigma)\uparrow\mbox{~as~}x_i\uparrow$. Further, if $R_i(\cdot)$ is a differential in $x_i$ $\forall i$, then,
    \begin{equation}
    	\label{Condition1:Eq}
		\frac{\partial R_i (\sigma)}{\partial x_i} > 0 \ \forall t_i .
	\end{equation}
	\end{condition}

\noindent\textit{Note.} If the strict inequality is replaced with $\geq$ in Eq.~\ref{Condition1:Eq}, we call it ``\textit{weak}" CM.
	
	\begin{condition}[Time Monotonicity]
		\label{Condition:2}
    The refund must always decrease with the increase in the duration of the project so as to incentivize early contribution i.e., $R(\sigma)$ must be a monotonically \emph{decreasing} function with respect to time $t_i \in (0,T),\forall x_i, \ \forall i \in N$ or 
    
    \begin{equation}\begin{split}
    \label{Condition2:Eq}
R_i(\sigma) \downarrow \mbox{ as } t_i \uparrow \mbox{ and } \exists \ t_i<T, \mbox{and }\Delta t_i \mbox{ s.t., }\\ \frac{ R_i\left((x_i,t_i+\Delta t_i),\sigma_{-i}\right) - R_i\left((x_i,t_i),\sigma_{-i}\right)}{\Delta t_i} < 0 
		\end{split}\end{equation}
	\end{condition}
	
Note that, with Condition \ref{Condition:2} we impose that $\not\exists t\in [0,T]$ such that there is a race among the agents to contribute at $t$.
We now analyze the consequence of such a refund bonus scheme on the game's characteristics induced by it.
    \subsection{Sufficiency of the Refund Bonus Scheme}
We show that a refund bonus scheme satisfying Conditions \ref{Condition:1} and \ref{Condition:2}, is sufficient to implement civic crowdfunding projects in sequential settings.
For this, let $G$ be the game induced by the refund bonus scheme $R(\cdot)$, for the payoff structure as given by Eq. \ref{GeneralPayoff}. We require $G$ to satisfy the following properties.
    \begin{Property}
    \label{Prop:1}
    In $G$, the total contribution equals the provision point at equilibrium, i.e., $C=H$. 
    \end{Property}

    \begin{Property}
    \label{Prop:2}
    $G$ must avoid the race condition.
    \end{Property}

    \begin{Property}
    \label{Prop:3}
    $G$ is a sequential game.
    \end{Property}

   \begin{customthm}{1S}
    \label{Theorem:General}
    Let $G$ be the game induced by a refund bonus scheme $R(\cdot)$ for the payoff structure as given by Eq. \ref{GeneralPayoff}, and with $\vartheta> H, 0<B<\vartheta-H$. If $R(\cdot)$ satisfies Conditions \ref{Condition:1} and \ref{Condition:2}, Properties \ref{Prop:1}, \ref{Prop:2} and \ref{Prop:3} hold.
	\end{customthm}
	\noindent\textit{Proof Sketch.}
	\begin{enumerate}[leftmargin=*]
	    \item Condition \ref{Condition:1} $\implies$ Property \ref{Prop:1}.	    At equilibrium, $C<H$ can not hold as $\exists i\in N$ with $x_i<\theta_i$, at least, since $\vartheta > H$. Such an Agent $i$ could obtain a higher refund bonus by marginally increasing its contribution since $R(\cdot)$ satisfies Condition \ref{Condition:1} and $B>0$. For $C>H$, any agent with a positive contribution could gain in payoff by marginally decreasing its contribution. 
	    \item Condition \ref{Condition:2} $\implies$ Properties \ref{Prop:2} and \ref{Prop:3}. Every Agent $i$ contributes as soon as it arrives, since $R(\cdot)$ satisfies Condition \ref{Condition:2}. This implies that, for the same contribution $x_i$ and for any $\epsilon>0$, we have $\pi_i(\cdot,y_i)>\pi_i(\cdot,y_i+\epsilon)$. Further, as the race condition is avoided, $G$ results in a sequential game. \qed
	\end{enumerate}

\subsection{Necessity of the Refund Bonus Schemes}
Theorem \ref{Theorem:General} shows that Condition \ref{Condition:1} is sufficient to satisfy Property \ref{Prop:1} and Condition \ref{Condition:2} is sufficient to satisfy Properties \ref{Prop:2} and \ref{Prop:3}. 
With Theorem~\ref{thm::neccGeneral}, we further prove that Condition~\ref{Condition:2} is necessary for Properties~\ref{Prop:2} and \ref{Prop:3}; while \textit{weak} Condition~\ref{Condition:1} is necessary for Property~\ref{Prop:1}. However, we remark that Theorem~\ref{thm::neccGeneral} does not characterize $G$ completely. For the theorem to hold, \textit{unlike} in the case of Theorem~\ref{Theorem:General}, we assume there exists a unique equilibrium defined by the strategy $(x_i^*,t_i^*),~\forall i\in N$.

\begin{customthm}{1N}\label{thm::neccGeneral}
    Let $G$ be the game induced by a refund bonus scheme $R(\cdot)$ for the payoff structure as given by Eq. \ref{GeneralPayoff}, and with $\vartheta> H, 0<B<\vartheta-H$. If $R(\cdot)$  satisfies Properties \ref{Prop:1}, \ref{Prop:2} and \ref{Prop:3} and there is unique equlibrium, then ``weak" Condition \ref{Condition:1} and Condition \ref{Condition:2} hold.
\end{customthm}
\noindent\textit{Proof Sketch.}
\begin{enumerate}[leftmargin=*]
    \item Property \ref{Prop:1} $\implies$ weak Condition~\ref{Condition:1}. Assume weak Condition~\ref{Condition:1} does not hold. This implies that $\exists i\in N$ for whom $R_i(x_i,\cdot)>R_i(x_i+\epsilon,\cdot)$ for some $\epsilon >0$. Now consider a case, wlog, that the agent $i$ is the last agent. Further, the project will be funded if agent $i$ contributes $x_i+\epsilon$, i.e., where its funded payoff equals its unfunded payoff\cite{zubrickas2014provision}. Since  $R_i(x_i,\cdot)>R_i(x_i+\epsilon,\cdot)$, agent $i$ will prefer to contribute $x_i$ and at equilibrium, $C\not = H$. This is a contradiction as it is given that Property~\ref{Prop:1} holds.
    \item Properties \ref{Prop:2} and \ref{Prop:3} $\implies$ Condition~\ref{Condition:2}. Property \ref{Prop:2} implies that $G$ avoids the race condition. That is, $\not\exists~i\in N$ for whom $\pi_i(x_i,y_i)>\pi_i(x_i,y_i+\epsilon)$ for any $\epsilon>0$ which in turn implies Condition~\ref{Condition:2}. This is because, for the same $x_i$, $\pi_i$ and $R_i$ are both decreasing with respect to $t_i$. \qed
    
\end{enumerate}

Theorem \ref{Theorem:General} shows that a refund bonus scheme satisfying Conditions \ref{Condition:1} and \ref{Condition:2} avoids the race condition (Property \ref{Prop:2}) and induces a sequential game (Property \ref{Prop:3}). Thus, a mechanism deploying such a refund bonus scheme can be \emph{implemented sequentially}, i.e., over web-based (or online) platforms.
Additionally, refund bonus schemes should also be clear to explain to a layperson. Moreover, these should be computationally efficient and cost-effective when deployed as a smart contract. 
Through this generalized result on refund bonus schemes, we show the following proposition.
    
	\begin{proposition}
    \label{prop:pps_c1c2}
    PPS satisfies Condition \ref{Condition:1} and Condition \ref{Condition:2}.
    \end{proposition}
	\begin{proof} Since every cost function used in PPS for crowdfunding must satisfy $\frac{\partial (r_i^{t_i}-x_i)}{\partial x_i} > 0$, $\forall i$ \cite[CONDITION-7]{chandra2016crowdfunding}, PPS satisfies Condition \ref{Condition:1}.
    
    For Condition \ref{Condition:2}, observe that $\forall i$, from \cite[Eq. 6]{chandra2016crowdfunding}
    	\begin{equation}\begin{split}
        	\label{PPS:Eq}
 				(r_i^{t_i}-x_i) & =  C_0^{-1}(x_i+C_0(q^{t_i}))-q^{t_i} -x_i. \\
		\end{split}\end{equation}
    
    In Eq. \ref{PPS:Eq}, as $t_i \uparrow$, $q^{t_i} \uparrow$ as it is a monotonically non-decreasing function of $t$ and thus R.H.S. of Eq. \ref{PPS:Eq} decreases since R.H.S. of Eq. \ref{PPS:Eq} is a monotonically decreasing function of $q^{t_i}$ \cite[Theorem 3 (Step 2)]{chandra2016crowdfunding}. Thus, PPS also satisfies Condition \ref{Condition:2}.\qed
    \end{proof}
    \begin{corollary}
    \label{cor:pps_race}
    PPS avoids the race condition and thus can be implemented sequentially.
    \end{corollary}

    In the following subsection, we present three novel refund schemes satisfying Conditions \ref{Condition:1} and \ref{Condition:2} and the novel provision point mechanisms based on them. 
 
    \subsection{Refund Bonus Schemes}
    \begin{table*}[!t]
        \centering
        \begin{adjustbox}{width=1\textwidth}
        \begin{tabular}{ccccl}
        \toprule
\textbf{Mechanism}  &       \textbf{Refund Scheme} & \textbf{Parameters} & \textbf{Covergence of Sum} & \textbf{Based On} \\
        \midrule
    PPRG &   $R^{PPRG}_i(\cdot)=\bigg(\frac{x_i+a \times (1/\gamma)^{i-1}}{C+K_1}\bigg)B$ & $a>0,1/\gamma<1,K_1=\frac{a\gamma}{\gamma-1}$ & $\sum_{i=1}^{\infty}\left(x_i+a(1/\gamma)^{i-1}\right)=C+K_1$ & Geometric Progression (GP)\\

    PPRE    & $R^{PPRE}_i(\cdot)=\bigg(\frac{x_i+K_2 \times e^{-t_i}}{C+K_2}\bigg)B $ & $K_2 > 0$ & $\sum_{i=1}^{\infty}(x_i)+\int_{t=t_1}^{\infty} (K_2 e^{-t} dt) \leq C+K_2$ &Exponential Function (EF) \\
  
    PPRP    & $R^{PPRP}_i(\cdot)=\bigg(\frac{x_i+K_3\times\frac{1}{i(i+1)} }{C+K_3}\bigg)B$ & $K_3 > 0$ & $\sum_{i=1}^{\infty}\left(x_i+K_3\frac{1}{i(i+1)}\right) = C+K_3$ &Polynomial Function (PF)\\
        \bottomrule
        
        \end{tabular}
        \end{adjustbox}
        \caption{Various Refund schemes satisfying Condition \ref{Condition:1} and Condition \ref{Condition:2} for an Agent $i$. Note that, in $R^{PPRG}\mbox{~and~}R^{PPRP}$, the subscript $i$ denotes the order of the contribution.}
        \label{tab:all_refund_schemes}
    \end{table*}

Table \ref{tab:all_refund_schemes} presents three novel refund schemes for an Agent $i$ contributing $x_i$ at time $t_i$ as well as the mechanisms which deploy them. Note that we require all the refund bonus schemes to converge to a particular sum that can be pre-computed. This convergence allows these schemes to be \emph{budget balanced}. The parameters $a,\gamma,K_1,K_2,K_3\mbox{~and~}B$ are mechanism parameters (for their respective mechanisms) which the SP is required to announce at the start.
Additionally, the refund schemes presented deploy three mathematical functions: geometrical, exponential, and polynomial decay. $R^{PPRG}(\cdot)$ and $R^{PPRP}(\cdot)$ refunds the contributing agents based on the sequence of their arrivals (similar to PPS), while the refund scheme $R^{PPRE}(\cdot)$ refunds them based on their time of contribution. 

\smallskip
\noindent\textit{Sufficiency Conditions.} We now show that PPRG satisfies Conditions \ref{Condition:1} and \ref{Condition:2}.

    \begin{Claim}
        \label{Claim:1}
   		$R^{PPRG}(\sigma)$ satisfies Condition \ref{Condition:1} $\forall i \in N$.
   		\end{Claim}
\begin{proof} Observe that $\forall i \in N$,
        		\begin{equation*}\begin{split}
 				\frac{\partial R^{PPRG}_i(\sigma)}{\partial x_i} & = \frac{B}{C+K_1} > 0 \ \forall t_i.\\
				\end{split}\end{equation*}
        
      Therefore, $R^{PPRG}(\cdot)$ satisfies Condition \ref{Condition:1} $\forall i$.\qed
     \end{proof}
    
		\begin{Claim}
        \label{Claim:2}
		$R^{PPRG}(\sigma)$ satisfies Condition \ref{Condition:2}.
		\end{Claim}
     \begin{proof} For every Agent $i \in N$ arriving at time $y_i$, its share of the refund bonus given by $R^{PPRG}(\cdot)$ will only decrease from that point in time, since its position in the sequence of contributing agents can only go down, making it liable for a lesser share of the bonus, for the same contribution. Let $\tilde{t_i}$ be the position of the agent arriving at time $y_i$, when it contributes at time $t_i$. While $\tilde{t_i}$ will take discrete values corresponding to the position of the agents, for the purpose of differentiation, let $\tilde{t_i} \in \mathbf{R}$. Now, we can argue that at every epoch of time $t_i$, Agent $\tilde{t_i}$ will contribute to the project. With this, $R^{PPRG}(\cdot)$ can be written as,
        	$$R^{PPRG}_i(\sigma)=\bigg(\frac{x_i+a \times (1/\gamma)^{\tilde{t_i}-1}}{C+K}\bigg)B.$$
      Further observe that $\forall i \in N$,
     		\begin{equation*} 
 				\frac{\partial R^{PPRG}_i(\sigma)}{\partial \tilde{t_i}} = -\bigg(\frac{a\times (1/\gamma)^{\tilde{t_i}}}{C+K_1}\bigg)B < 0 \ \forall x_i. 
				 \end{equation*}
        Therefore, $R^{PPRG}(\cdot)$ satisfies Condition \ref{Condition:2}.\qed
        \end{proof}


We can similarly prove that $R^{PPRE}\mbox{~and~}R^{PPRP}$ satisfy Conditions \ref{Condition:1} and \ref{Condition:2}.

    \subsection{Gas Comparisons}
        \label{Simulation:SmartContract}
As aforementioned, CC is now being deployed
as smart contracts (SCs) over the Ethereum network. Thus, CC mechanisms deployed as SCs must be efficient, i.e., result in less \textit{gas} consumption. Gas is a unit of fees that the Ethereum protocol charges per computational step executed in a contract or transaction. This fee prevents deliberate attacks and abuse on the Ethereum network \cite{buterin14ethereum}.

We show a hypothetical cost comparison between PPS, PPRG, PPRE, and PPRP based on the Gas usage statistics from \cite{buterin14ethereum,wood2014ethereum}. For the relevant operations, the cost in Gas units is: ADD: 3, SUB: 3, MUL: 5, DIV: 5, EXP($x$): $10+10*log(x)$ and LOG($x$): $365+8*\mbox{size of $x$ in bytes}$. Table \ref{Table-Supplement:Gas} presents the comparison\footnote{We do not require any exponential calculation in PPRG -- by storing the last GP term in a temporary variable.}. We remark that the only difference in the induced CC game will be the computation of the refund bonus for each contributing agent. This refund will depend on the underlying refund bonus scheme. Thus, we focus only on the gas cost because of the said schemes.

From Table~\ref{Table-Supplement:Gas}, for every agent, PPRG takes $21$ gas units, PPRP takes $31$ gas units, PPRE takes at least $31$ gas units, and PPS takes at least $407$ gas units. When implemented on smart contracts, PPS is an expensive mechanism because of its logarithmic scoring rule for calculating payment rewards. PPRG, PPRP, and PPRE, on the other hand, use simpler operations and therefore have minimal operational costs.

\smallskip
\noindent\textit{Inference from Table~\ref{Table-Supplement:Gas}.} Note that the average gas price per unit varies.  At the time of writing this paper, we have the average gas price $\approx 200$ GWei, i.e., $2\times 10^{-7}$ ETH; and also $1\mbox{~ETH~}\approx 1162$ USD. As a result, the cost incurred by a crowdfunding platform, assuming when $n=100$, is (approximately) (i) PPS: 10 USD (at least); (ii) PPRG: 0.5 USD; (iii) PPRE: 0.72 USD (at least); and (iv) PPRP: 0.72 USD.  Further, in December 2019, Kickstarter had $3524$ active projects~\cite{wiki:kickstarter}. The data implies the total cost across the projects for (i) PPS: 35240 USD; and (ii) PPRG: 2537.28 USD. PPRG reduces the cost incurred by the platform by (at least) $\approx 14$ times.

\begin{table*}[!t]
\centering
\begin{adjustbox}{width=1\textwidth}
\begin{tabular}{ccccccccc}
\toprule
  \multirow{2}{*}{\textbf{Operation}} 
      & \multicolumn{2}{c}{\textbf{PPS}} 
          & \multicolumn{2}{c}{\textbf{PPRG}}  & \multicolumn{2}{c}{\textbf{PPRE}} &
          \multicolumn{2}{c}{\textbf{PPRP}} \\
  & Operations & Gas Consumed &Operations & Gas Consumed & Operations & Gas Consumed & Operations & Gas Consumed \\  
  \midrule
  ADD& 2 & 6 & 2 & 6 & 2 & 6 & 2 & 6 \\     
  SUB& 2 & 6 & 0 & 0 & 0 & 0 & 0 & 0 \\     
  MUL& 2 & 10 & 2 & 10 & 2 & 10 & 3 & 15 \\     
  DIV& 2 & 10 & 1 & 5 & 1 & 5 & 2 & 10 \\      
  EXP($x$) & 2  & $10+10\times(log(x))$ & 0 & 0 & 1 &  $10+10\times(log(x))$ & 0 & 0 \\    
  LOG($x$)& 2 & $365+8\times(\mbox{bytes logged})$ & 0 & 0 & 0 &0 & 0 & 0 \\     
  
  & \textbf{Total Gas:} &407 (at least) & \textbf{Total Gas:} & 21 & \textbf{Total Gas:} & 31 (at least) & \textbf{Total Gas:} & 31 \\ 
  \bottomrule
\end{tabular}
\end{adjustbox}
\caption{\label{Table-Supplement:Gas}Gas Consumption comparison between PPS, PPRG, PPRE and PPRP for an agent. All values are in Gas units.}

\normalsize
\end{table*}

\section{PPRG\label{sec::PPRG}}

	We now describe the mechanism \emph{Provision Point mechanism with Refund through Geometric Progression} (PPRG), for crowdfunding a public project. PPRG incentivizes an interested agent to contribute as soon as it arrives at the crowdfunding platform. In PPRG, for the exact contribution of Agent $i$ and Agent $j$, i.e., $x_i=x_j$, the one who contributed earlier obtains a higher share of the refund bonus. These differences in shares are allocated using an infinite geometric progression series (GP) with a common ratio of $< 1$. 
    
    \smallskip
   \noindent{\textit{Refund Bonus Scheme.}}
        The sum of an infinite GP with $a>0$ as the first term and $0<1/\gamma<1$ as the common ratio is: $K_1=a\times \sum_{i=0}^{\infty}(1/\gamma)^i=\frac{a\gamma}{\gamma-1}.$ With this, we propose a novel refund bonus scheme,  \begin{equation}\label{Bonus:PPRG}R^{PPRG}_i(\sigma)=p_i=\bigg(\frac{x_i+a \times (1/\gamma)^{i-1}}{C+K_1}\bigg)B \end{equation} for every Agent $i\in N$, $B>0$ as the total bonus budget allocated for the project by the SP and where $\sigma=\left((x_i,t_i) \: | \: \forall i \in N \right)$. The values $a$ and $\gamma$ are mechanism parameters which the SP is required to announce at the start of the project. 
  
  \smallskip
\noindent  \textit{Equilibrium Analysis of PPRG.} The analysis follows from Theorem~\ref{Theorem:General}.
        	\begin{customthm}{2}
            \label{Theorem:PPRG}
        	For PPRG, with the refund $p_i$ as described by Eq. \ref{Bonus:PPRG} $\forall i \in N$, satisfying $0 < B\leq \vartheta-H$ and with the payoff structure as given by Eq. \ref{GeneralPayoff}, a set of strategies $\bigg\{\left(\sigma^*_i=(x_i^*,y_i)\right):\ if \ h^{y_i} =0 \ then \ x^*_i=0 \ otherwise \ x_i^* \leq \frac{\theta_i(H+K_1)-aB\times (1/\gamma)^{i-1}}{H+K_1+B} \bigg\} \: \forall i \in N$ are sub-game perfect equilibria, such that at equilibrium $C=H$. In this, $x^*_i$ is the contribution towards the project, $y_i$ is the arrival time to the project of Agent $i$, respectively.
        	\end{customthm}
\begin{proof} We prove the theorem with the following steps.

\smallskip
  \noindent\underline{\emph{Step 1}:} Since $R^{PPRG}(\cdot)$ satisfies Condition \ref{Condition:1}  (Claim \ref{Claim:1}) and Condition \ref{Condition:2} (Claim \ref{Claim:2}) and has a payoff structure as given by Eq. \ref{GeneralPayoff}, from  Theorem \ref{Theorem:General} we get the result that PPRG induces a sequential move game and thus, can be implemented in a sequential setting. 
  
  \smallskip
            \noindent \underline{\emph{Step 2}:}  From Claim 2, the best response for any agent is to contribute as soon as he arrives i.e., at time $y_i$.
             
             \smallskip
           \noindent \underline{\emph{Step 3}:}  We assume that each agent is symmetric in its belief for the provision of the project. Moreover, from Theorem \ref{Theorem:General}, agents know that the project will be provisioned at equilibrium. Therefore, for any agent, its equilibrium contribution becomes that $x_i^*$ for which its provisioned payoff is \emph{greater than or equal to} its not provisioned payoff.
            Now, with $C=H$ at equilibrium,
        $$\theta_i - x_i^* \geq p_i = {\bigg(\frac{x_i^*+a \times (1/\gamma)^{i-1}}{C+K_1}\bigg)B} $$
        $$\Rightarrow x_i^*\leq \frac{\theta_i(H+K_1)-aB\times (1/\gamma)^{i-1}}{H+K_1+B}$$
       	
       	\smallskip
        \underline{\emph{Step 4}:}  Summing over $x_i^*,~\forall i$ we get,
                \begin{equation*}B\leq \frac{(H+K_1)\vartheta-H^2-HK_1}{H+K_1}.\end{equation*} 
                
               as $\sum_{i\in N} x_i^* = H$. From the above equation, we get
                \begin{equation*}
                \begin{aligned}
                &0 < B\leq \frac{(H+K_1)\vartheta-H^2-HK_1}{H+K_1}=\vartheta - H
                \end{aligned}
                \end{equation*} as a sufficient condition for existence of Nash Equilibrium for PPRG.

       	\smallskip
       	\noindent	\underline{\emph{Step 5}:} The following scenarios prove that the strategies are sub-game perfect.
       	    \begin{itemize}[leftmargin=*]
       	        \item For an Agent $i$ entering the project such that $h^{y_i}=0$ (i.e., $C=H$), its best response is contributing $0$.
       	        \smallskip
       	        \item For an Agent $i$ entering the project such that $h^{y_i}>0$ with $x_i^* > h^{y_i}$, its  best response is contributing $h^{y_i}$. Observe that, Agent $i$ will contribute the maximum contribution required, $h^{y_i}$, since its not provisioned payoff increases as its contribution increases (Claim \ref{Claim:1}). Therefore, for a contribution less than $h^{y_i}$, Agent $i$ will receive \emph{lesser} payoff in comparison for the contribution $h^{y_i}$.  
       	        \smallskip
    		    \item Lastly, for an Agent $i$ entering the project such that $h^{y_i}>0$ with $x_i^* \leq h^{y_i}$, its best response is contributing $x_i^*$ (as defined in Theorem \ref{Theorem:PPRG}). This is because for the contribution $x_i^*$, its provisioned payoff is \emph{equal} to its not provisioned payoff. For this scenario, with backward induction, it is the best response for every Agent $i$ to follow the same strategy in which their provisioned payoffs are equal to their not provisioned payoffs, irrespective of $h^{y_i}$.\qed
       	    \end{itemize}
    \end{proof}
  
\noindent\textit{Discussion.} Observe that, as the refund bonus decreases with time (Claim \ref{Claim:2}), each agent in PPRG is better off contributing once instead of breaking up its contribution. This result follows as we assume that each agent's belief for the project's provision is symmetric and does not vary.

With Theorem \ref{Theorem:PPRG}, we identify a set of pure-SPE at which the project is provisioned. However, we do not claim that these are the only set of pure-SPE possible. We leave it for future work to explore other possible pure-SPE at which the project gets provisioned. Also, the equilibrium analysis of PPRE and PPRP is similar to Theorem \ref{Theorem:PPRG}.

\smallskip
\noindent\textit{Coalition-proof.} Along similar lines of the argument presented in~\cite[Section 4.2]{zubrickas2014provision}, we can show that the game induced in PPRG will be coalition-proof. This is because the equilibrium in the induced game follows the \textit{aggregate concurrence principle}~\cite{stole}, i.e., at equilibrium, agents must agree on the choice of aggregate outcomes. As it immediately follows from this principle, the equilibria produced by PPRG (Theorem~\ref{Theorem:PPRG}) are coalition-proof.

\section{Conclusion}
In this paper, we looked for provision point mechanisms for CC with refund bonus schemes. Towards it, we introduced Contribution Monotonicity and Time Monotonicity for refund bonus schemes in CC mechanisms. We proved that these two conditions are sufficient to implement provision point mechanisms with refund bonuses to possess an equilibrium that avoids free-riding and the race condition (Theorem \ref{Theorem:General}).  We then proposed three simple refund bonus schemes and design novel mechanisms that deploy them, namely, PPRG, PPRE, and PPRP. We showed that PPRG has much less cost when implemented as a smart contract over the Ethereum framework. We identified a set of sub-game perfect equilibria for PPRG in which it provisions the project at equilibrium (Theorem \ref{Theorem:PPRG}).


  \printbibliography

\newpage  
 \appendix

\section{Proof of Theorem 1S}

In Steps 1, 2 and 3, we show that $R(\cdot)$ satisfying Condition \ref{Condition:1} is sufficient to satisfy Property \ref{Prop:1} and Condition \ref{Condition:2} is sufficient to satisfy Properties \ref{Prop:2} and \ref{Prop:3}.
    		\squishlist
    			\item \underline{\textit{Step 1}}: As $\vartheta> H$, from Eq. \ref{GeneralPayoff}, at equilibrium $C<H$ cannot hold, as $\exists i\in N$ with $x_i<\theta_i$, at least. Such an Agent $i$ could obtain a higher refund bonus by marginally increasing its contribution since $R(\cdot)$ satisfies Condition \ref{Condition:1} and $B>0$. For $C>H$, any agent with a positive contribution could gain in payoff by marginally decreasing its contribution. Thus, at equilibrium $C=H$ or $G$ satisfies Property \ref{Prop:1}.
    		
    		\item \underline{\textit{Step 2}}:  Every Agent $i$ contributes as soon as it arrives, since $R(\cdot)$ satisfies Condition \ref{Condition:2} i.e., $\forall i \in N$, $$\pi_i\left((x_i,y_i),\sigma_{-i}\right)>\pi_i\left((x_i,t),\sigma_{-i}\right)\: \forall t \in (y_i,T].$$ In other words, the best response $\forall i\in N$ is the strategy $\sigma_i=(x_i,y_i)$. Thus, as per Definition \ref{def:race_cond}, $G$ avoids the race condition or $G$ satisfies Property \ref{Prop:2}.

    		\item \underline{\textit{Step 3}}: Since $G$ satisfies Property \ref{Prop:2}, it avoids the race condition. Hence, it can be implemented in a sequential setting or $G$ is a sequential game. \qed

    		\squishend

\section{\label{app:sec:rbs}Refund Bonus Schemes}

        \begin{claim}
    \label{Claim:1:PPRE}
   		$R^{PPRE}(\sigma)$ satisfies Condition 1 $\forall i \in N$.
   		\end{claim}
     \noindent   \textbf{Proof:} Observe that $\forall i \in N$,
        		\begin{equation*}\begin{split}
 				\frac{\partial R^{PPRE}_i(\sigma)}{\partial x_i} & = \frac{B}{C+K_2} > 0 \ \forall t_i.\\
				\end{split}\end{equation*}
        
      Therefore, $R^{PPRE}(\cdot)$ satisfies Condition 1 $\forall i$. \qed
    
     \begin{claim}
    \label{Claim:2:PPRE}
   		$R^{PPRE}(\sigma)$ satisfies Condition 2 $\forall i \in N$.
   		\end{claim}
   \noindent     \textbf{Proof:} Observe that $\forall i \in N$,
        		\begin{equation*}\begin{split}
 				\frac{\partial R^{PPRE}_i(\sigma)}{\partial t_i} & = -\left(\frac{K_2B}{C+K_2}\right) < 0 \ \forall x_i.\\
				\end{split}\end{equation*}
        
      Therefore, $R^{PPRE}(\cdot)$ satisfies Condition 2 $\forall i$. \qed
  
            \begin{claim}
    \label{Claim:1:PPRP}
   		$R^{PPRP}(\sigma)$ satisfies Condition 1 $\forall i \in N$.
   		\end{claim}
    \noindent    \textbf{Proof:} Observe that $\forall i \in N$,
        		\begin{equation*}\begin{split}
 				\frac{\partial R^{PPRP}_i(\sigma)}{\partial x_i} & = \frac{B}{C+K_3} > 0 \ \forall t_i.\\
				\end{split}\end{equation*}
        
      Therefore, $R^{PPRP}(\cdot)$ satisfies Condition 1 $\forall i$. \qed
    
    \begin{claim}
        \label{Claim:2:PPRP}
		$R^{PPRP}(\sigma)$ satisfies Condition 2.
		\end{claim}
     \noindent   \textbf{Proof:}  For every Agent $i \in N$ arriving at time $y_i$, its share of the refund bonus given by $R^{PPRP}(\cdot)$ will only decrease from that point in time, since its position in the sequence of contributing agents can only go down, making it liable for a lesser share of the bonus, for the same contribution. Let $\tilde{t_i}$ be the position of the agent arriving at time $y_i$, when it contributes at time $t_i$. While $\tilde{t_i}$ will take discrete values corresponding to the position of the agents, for the purpose of differentiation, let $\tilde{t_i} \in \mathbf{R}$. Now, we can argue that at every epoch of time $t_i$, Agent $\tilde{t_i}$ will contribute to the project. With this, $R^{PPRP}(\cdot)$ can be written as,
        	$$R^{PPRP}_i(\sigma)=\left(\frac{x_i+K_3\times\frac{1}{\tilde{t_i}(\tilde{t_i}+1)} }{C+K_3}\right)B.$$
      Further observe that $\forall i \in N$,
     		\begin{equation*} 
 				\frac{\partial R^{PPRP}_i(\sigma)}{\partial \tilde{t_i}} = \frac{K_3B}{C+K_3}\left(-\frac{1}{\tilde{t_i^2}}+\frac{1}{(\tilde{t_i}+1)^2}\right) < 0 \ \forall x_i. 
				 \end{equation*}
        The inequality follows from the fact that $\frac{1}{\tilde{t_i^2}}>\frac{1}{(\tilde{t_i}+1)^2}$ as $\tilde{t_i}>0$. Therefore, $R^{PPRG}(\cdot)$ satisfies Condition 2. \qed

\section{Equilibrium Analysis of PPRE} \label{app:ppre}
\begin{theorem}
            \label{Theorem:PPRE}
        	For PPRE, with the refund $p_i$ as described in Table 1 (in the paper) $\forall i \in N$, $\vartheta\geq H$ $C=H$, which satisfies $0 < B\leq \vartheta-H$ and has the payoff structure as given by Eq. 1, the set of strategies $\bigg\{\left(\sigma^*_i=(x_i^*,y_i)\right):\ if \ h^{y_i}=0 \ then \ x^*_i=0 \ otherwise \ x_i^* \leq \frac{\theta_i(H+K_2)-BK_2\times e^{-y_i}}{H+K_2+B} \bigg\} \: \forall i \in N$ are sub-game perfect equilibria. In this, $x^*_i$ is the contribution towards the project, $y_i$ is the arrival time to the project of Agent $i$, respectively.
\end{theorem}
\noindent \textbf{Proof.} The proof for the theorem follows similar to as presented for Theorem 2. The condition for the existence of Nash Equilibrium for PPRE is given as,
    \begin{equation*}
                \begin{aligned}
                &0 < B\leq \frac{(H+K_2)\vartheta-H^2-HK_2}{H+K_2}\\
                \implies&0 < B \leq \vartheta - H.
                \end{aligned}
                \end{equation*} \qed

\begin{figure*}[!t]
    \centering
        \subfloat[]{\includegraphics[width=0.45\linewidth]{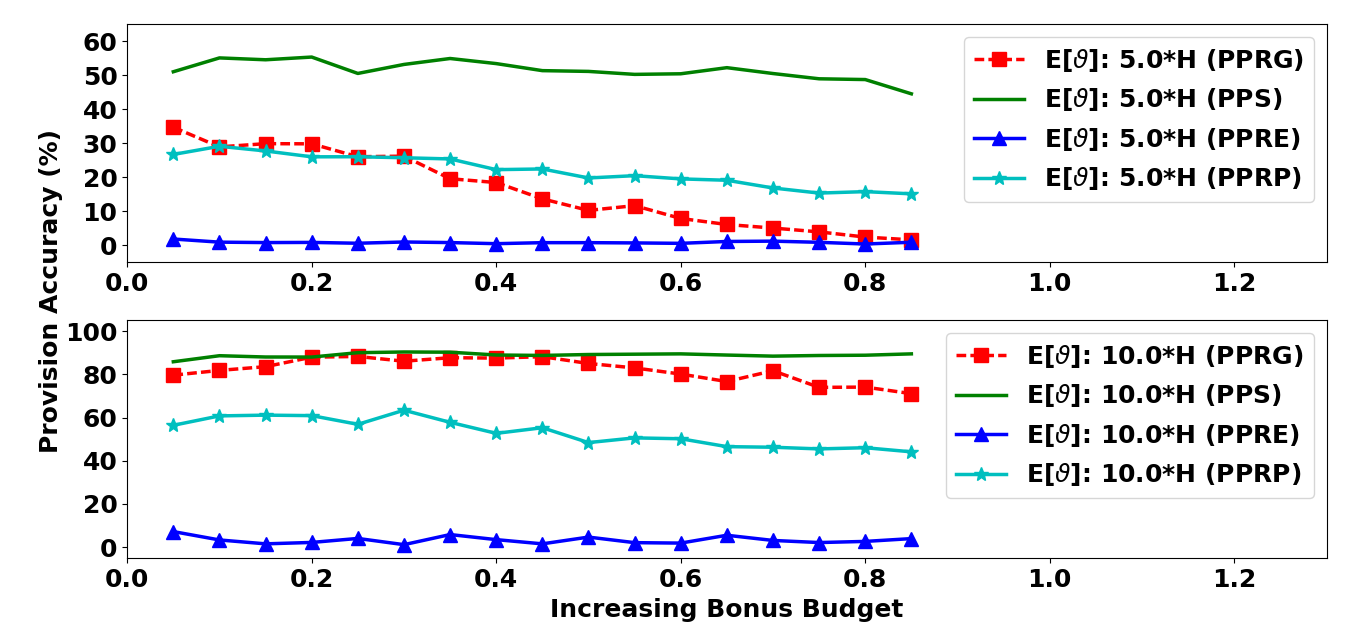}}
        \qquad
        \subfloat[]{\includegraphics[width=0.45\linewidth]{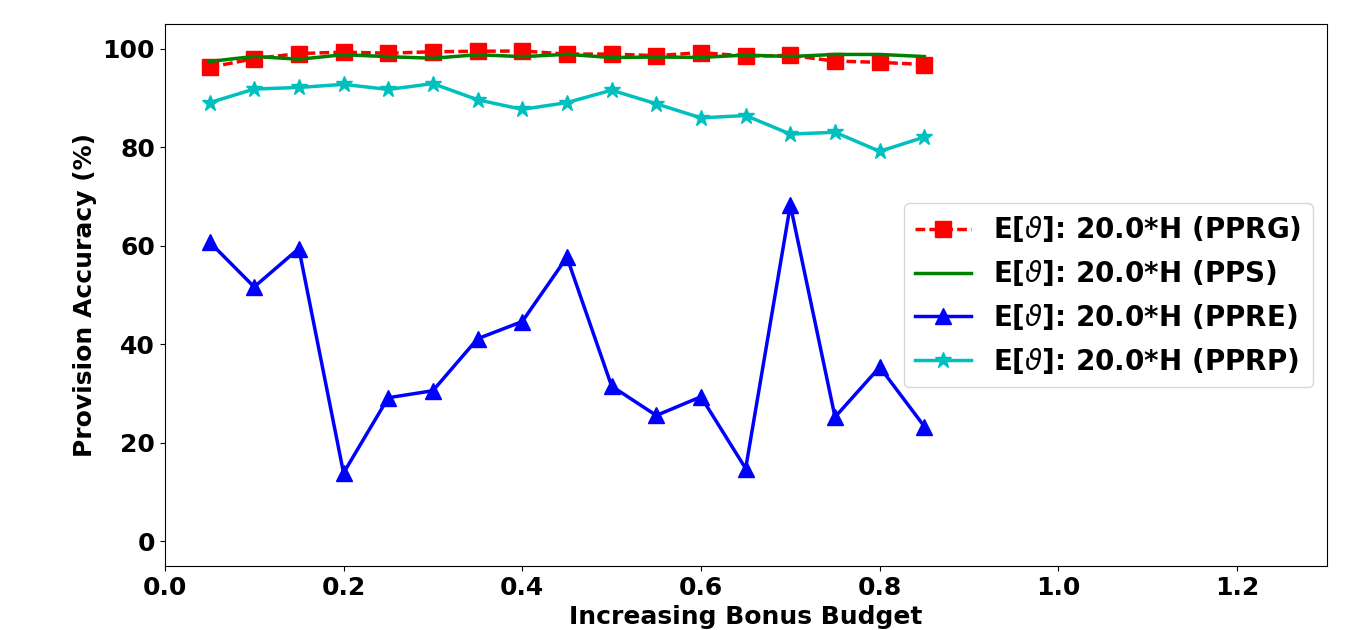}}
            \caption{Comparison of provision accuracy of PPRG, PPRE and PPRP with PPS for (a) $E[\vartheta]=5*H$ (top) (a) $E[\vartheta]=10*H$ (bottom) and (b) $E[\vartheta]=20*H$.}
    \label{fig:pprg-pps-n-25-Var-20}
\end{figure*}

\section{Equilibrium Analysis of PPRP}\label{app:pprp}

\begin{theorem}
            \label{Theorem:PPRP}
        	For PPRP, with the refund as described in Table 1 (in the paper) $\forall i \in N$, $\vartheta\geq H$ $C=H$, which satisfies $0 < B\leq \vartheta-H$ and has the payoff structure as given by Eq. 1, the set of strategies $\bigg\{\left(\sigma^*_i=(x_i^*,y_i)\right):\ if \ h^{y_i}=0 \ then \ x^*_i=0 \ otherwise \ x_i^* \leq \frac{\theta_i(H+K_3)-BK_3\times\frac{1}{i(i+1)} }{H+K_3+B} \bigg\} \: \forall i \in N$ are sub-game perfect equilibria. In this, $x^*_i$ is the contribution towards the project, $y_i$ is the arrival time to the project of Player $i$, respectively.
\end{theorem}
\noindent \textbf{Proof.} The proof for the theorem follows similar to as presented for Theorem 2. The condition for the existence of Nash Equilibrium for PPRP is given as,
    \begin{equation*}
                \begin{aligned}
                &0 < B\leq \frac{(H+K_3)\vartheta-H^2-HK_3}{H+K_3}\\
                \implies&0 < B \leq \vartheta - H.
                \end{aligned}
                \end{equation*} \qed

\section{Simulation Analysis}
\label{Simulation:RL}
In Section \ref{Simulation:SmartContract} we analyzed PPRG, PPRE, and PPRP in a hypothetical cost comparison with respect to PPS if they were implemented as smart contracts. In this section, we compare PPRG, PPRE, PPRP, and PPS for provision accuracy using a civic crowdfunding proprietary simulator built in partnership with \emph{KoineArth}~\cite{koinearth}.

However, it is very challenging to test civic crowdfunding mechanisms in a real-world environment because of the irreversible nature of the civic properties and decisions made in the process. Therefore, we employ Reinforcement Learning (RL) based simulations to test and compare the applicability and performance of the mechanisms. RL is an area of machine learning where agents interact with an environment and learn through a trial and error process where each action is rewarded or penalized based on its consequences on the game.

In this simulator, we create a Reinforcement Learning environment for PPRG, PPRE, PPRP, and PPS where agents learn to participate in the mechanisms. Agents go through repetitive iterations and learn their best strategy through rewards distributed by the corresponding mechanism. We run the simulation of 25 agents for all the mechanisms and obtain comparison results between PPRG, PPRE, PPRP with respect to PPS. In order to measure the performance of these mechanisms, we define the quantity \emph{provision accuracy}. For a mechanism $\mathcal{M}$, the provision accuracy is defined as the fraction of the civic projects provisioned by $\mathcal{M}$ over the total number of projects simulated. The results of the simulation are shown in Figure \ref{fig:pprg-pps-n-25-Var-20}.

Among PPRG, PPRE, and PPRP, it is clear to see that PPRG shows better provision accuracy. 
In case when the \emph{total expected valuation} ($E(\vartheta)$) is low (5 times the provision point), PPRP shows slightly better accuracy. However, the gain in the accuracy only comes at the expense of a budget very close to the maximum possible budget, i.e., $B=E(\vartheta)-H$. Such a budget is difficult to get in realistic circumstances. 
Note that, the equilibrium contributions are such that the provisioned payoff equals the not provisioned payoff (as defined in Theorem \ref{Theorem:PPRG}). Therefore, the difference in the accuracy can be attributed to the greater refund share provided by PPRG, for the same budget. This increases the not provisioned payoff for the agents, thereby incentivizing them to increase their contributions.
Thus, we conclude that PPRG performs better than PPRE and PPRP. 

When compared to PPS, PPRG shows significantly good provision accuracy when $E(\vartheta)$ is high (10 times provision point, for instance). When PPS shows a slightly higher accuracy, it again comes at the expense of a budget close to the maximum possible budget, $B$. For a reasonable budget of approximately $0.5 \times B$ or less, both the mechanisms share similar accuracy. Thus, PPS and PPRG perform equally in terms of provision accuracy, for a rational budget.

\end{document}